# Are Shockley-Read-Hall and ABC models valid for lead halide perovskites?

Alexander Kiligaridis,[1] Pavel Frantsuzov[*,2] Aymen Yangui,[1] Sudipta Seth,[1] Jun Li,[1] Qingzhi An,[3] Yana Vaynzof[*3] and Ivan G. Scheblykin[*1]

[1]*Chemical Physics and NanoLund, Lund University, P.O. Box 124, 22100 Lund, Sweden*

[2]*Voevodsky Institute of Chemical Kinetics and Combustion, SB RAS, Novosibirsk, Russia*

[3]*Integrated Center for Applied Physics and Photonic Materials (IAPP) and Centre for Advancing Electronics Dresden (CFAED), Technical University of Dresden, Dresden, Germany*

[*] **Corresponding Authors:**

ivan.scheblykin@chemphys.lu.se

yana.vaynzof@tu-dresden.de

pavel.frantsuzov@gmail.com

**TOC figure**

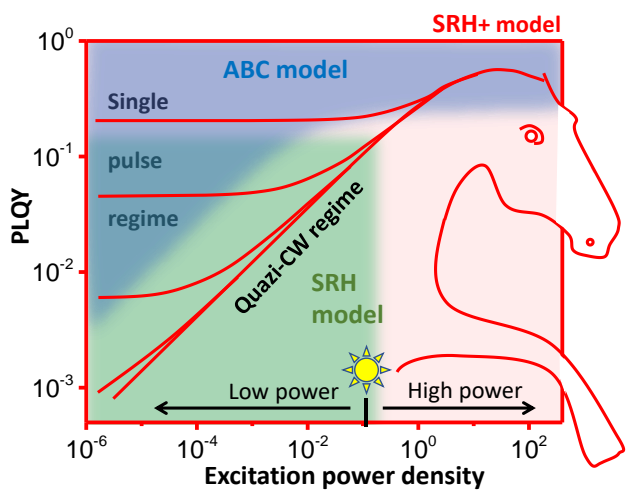

**Abstract**

Metal halide perovskites are an important class of emerging semiconductors. Their charge dynamics is poorly understood due to limited knowledge of defect physics and charge recombination mechanisms. Nevertheless, classical ABC and Shockley-Read-Hall (SRH) models are ubiquitously applied to perovskites without considering their validity. Herein, an advanced technique mapping photoluminescence quantum yield (PLQY) as a function of both the excitation pulse energy and repetition frequency is developed and employed to examine the validity of these models. While ABC and SRH fail to explain the charge dynamics in a broad range of conditions, the addition of Auger recombination and trapping to the SRH model enables a quantitative fitting of PLQY maps and low-power PL decay kinetics, and extracting trap concentrations and efficacies. Higher-power PL kinetics requires the inclusion of additional non-linear processes. The PLQY mapping developed herein is suitable for a comprehensive testing of theories and is applicable to any semiconductor.

## 1. Introduction

Semiconducting materials often exhibit a complex charge dynamics, which strongly depends on the concentration of charge carriers due to the co-existence of both linear and non-linear charge recombination mechanisms.[1,2] The emergence of novel semiconductors like metal halide perovskites exhibiting intriguing and often unexpected electronic properties, [3–10] triggered a renewed interest in revisiting the classical textbook theories of charge recombination and the development of more complete, accurate models.[11–17] Moreover, modern technical advances in experimental and computational capabilities[4,18–21] allow for a detailed quantitative comparison between experiment and theory, far beyond what was once possible.

Metal halide perovskites (MHP) are a novel solution-processable material class with enormous promise for application in a broad range of optoelectronic devices.[22–24] Driven in particular by their remarkable performance in photovoltaics, with power conversion efficiencies surpassing 25% demonstrated to date,[25] significant research efforts have been devoted to studying the fundamental electronic properties of these materials. [4,5,7,13,15,18,26–30] It was established that for many MHP compositions - with the most notable example being the methylammonium lead triiodide (MA=$CH_3NH_3^+$, also referred as $MAPbI_3$ or MAPI) - they can be considered as classical crystalline semiconductors at room temperature, in which photoexcitation leads to the formation of charge carriers that exist independently from each other due to the low exciton binding energy.[26] Consequently, conventional models that describe the charge carrier dynamics are ubiquitously used to describe the dynamics of charge carriers in MHPs.[11,13–15,31–38]

Historically, the first model describing the kinetics of charge carrier concentrations in a semiconductor was proposed by Shockley and Read [39] and independently by Hall [40], and is known as the Shockley-Read-Hall (SRH) model. This model considers only the first order process (trapping of electrons or holes) and the second order kinetic processes (radiative electron-hole recombination and non-radiative (NR) recombination of the trapped electrons and free holes). It is noteworthy that the SRH model allows the concentrations of free charge carriers to differ due to the presence of trapping. In an intrinsic semiconductor, trapping of, for example, electrons generated by photoexcitation creates excess of free holes at the valence band. This effect is often referred to as photodoping, in analogy with chemical doping, with the important difference, however, that the material becomes doped only under light irradiation and the degree of doping depends on the light irradiation intensity.

Third order processes, such as non-radiative Auger recombination, *via* which two charge carriers recombine in the presence of third charge that uptakes the released energy, have been recognised as particularly important at a high charge carrier concentration regime. To account for this process, Shen *et al.*, instead of adding the Auger recombination term into the SRH model, proposed a simplified ABC model named after the coefficients A, B and C for the first order (monomolecular), second order (bimolecular) and third order Auger recombination, respectively.[41] These coefficients are also sometimes referred to as $k_1$, $k_2$ and $k_3$. Importantly, the concentrations of free electrons and holes in the ABC model are assumed to be equal, thus neglecting the possible influences of chemical and photodoping effects. The ABC model is widely applied in a broad range of semiconductors and in particular, is commonly used to rationalize properties and efficiency limits of LEDs [41,42] The simplicity of the ABC model led to its extreme popularity also for MHPs (see e.g. [15] and references there in) with fewer reports employing SRH or its modifications.[13,14,16,17,33,38,43]

The ABC and SRH kinetic models are typically employed to describe experimentally acquired data such as the excitation power density dependence of PL quantum yield (PLQY) measured upon continues wave (CW) or pulsed excitation, time-resolved PL decay kinetics and kinetics of the transient absorption signal. These models are applied to semi-quantitatively explain the experimental results and extract different rate constants, [13–15,20,33–36,44,45] often without necessarily considering the models' limitations. Despite the very large number of published studies describing electronic processes in MHPs using the terminology of classical semiconductor physics to the best of our knowledge, there have been only very few attempts to fit both PL decay and PLQY dependencies of excitation power using ABC/SRH-based models,[14,16,17,33] These attempts, however, were of limited success because large discrepancies between the experimental results and the theoretical fits were often permitted.

These observations raise fundamental questions concerning the general validity of the SRH and ABC models to MHPs and the existence of a straight-forward experimental method to evaluate this validity. To address these concerns, it is necessary to characterise experimentally the PLQY and PL decay dynamics not only across a large range of excitation power densities, but also simultaneously over a large range of the repetition rates of the laser pulses. We note PL is sensitive not only to the concentrations of free charge carriers, but also, indirectly, to the concentration of trapped charge carriers, as the latter influence the former via charge neutrality. Such trapped carriers may also lead to other non-linear processes, for example, between free and trapped charge carriers (Auger trapping[2]), which should also be considered, but are not included in neither the classical SRH nor the ABC models. To expose and probe these processes, it is most crucial to scan the laser repetition rate frequency in the PLQY measurements, with such measurements, to the best of our knowledge, have not been reported to date.

In this work, we developed a new experimental methodology that maps the PLQY as a function of both the excitation pulse fluence (P, in photons/cm$^2$) and excitation frequency (f, in Hz). The novel technique allows to unambiguously determine the excitation regime of the sample (single pulse *vs* quasi – CW), which is critically important for data interpretation and modelling. By applying this method to a series of MAPbI$_3$ samples, we demonstrate that neither ABC nor classical SRH model can fit the acquired PLQY maps across the entire excitation parameter space. To tackle this issue, we develop an enhanced SRH model (in the following, the SRH+ model), which accounts for Auger recombination and Auger trapping processes and demonstrates that SRH+ is able to describe and quantitatively fit the PLQY(f,P) map over the entire range of excitation conditions with excellent accuracy. PL decays can be also fitted, albeit, with a more moderate accuracy. The application of the SRH+ model allowed us to extract the concentration of dominant traps in high electronic quality MAPbI$_3$ films to be of the order of $10^{15}$ cm$^3$ and to demonstrate that surface treatments can create a different type of trapping states of much higher concentration. Beyond the quantitative success of the extended SRH+ model, we reveal that there are indications of the presence of further non-linear mechanisms that influence charge dynamics at high charge carrier concentrations in MHPs.

## 2. PLQY(f,P) mapping and elucidation of the excitation regime

The acquisition of a PLQY(f,P) map occurs by measuring the intensity of PL for a series of pulse fluences (P) for each of which the pulse frequency (f) is scanned across a broad range (Fig. 1 a, c). Further details are provided in Supplementary Notes 1-3. In our case, the frequency is scanned from 100 Hz to 80 MHz, which corresponds to a lag between pulses varying from 12.5 ns to 10 ms. After scanning the frequency for a certain value of P, it is then changed to the next value and the scanning procedure is repeated. The pulse fluence ranges over 4 orders of magnitude (**P1**=4.1x10$^8$, **P2**=4.9x10$^9$, **P3**=5.1x10$^{10}$, **P4**=5.5x10$^{11}$ and **P5**=4.9x10$^{12}$ photons/cm$^2$). Such fluences, in the single pulse regime (see below), correspond to charge carrier densities of 1.04x10$^{13}$, 1.24x10$^{14}$, 1.3x10$^{15}$, 1.37x10$^{16}$ and 1.24x10$^{17}$ cm$^{-3}$, respectively.

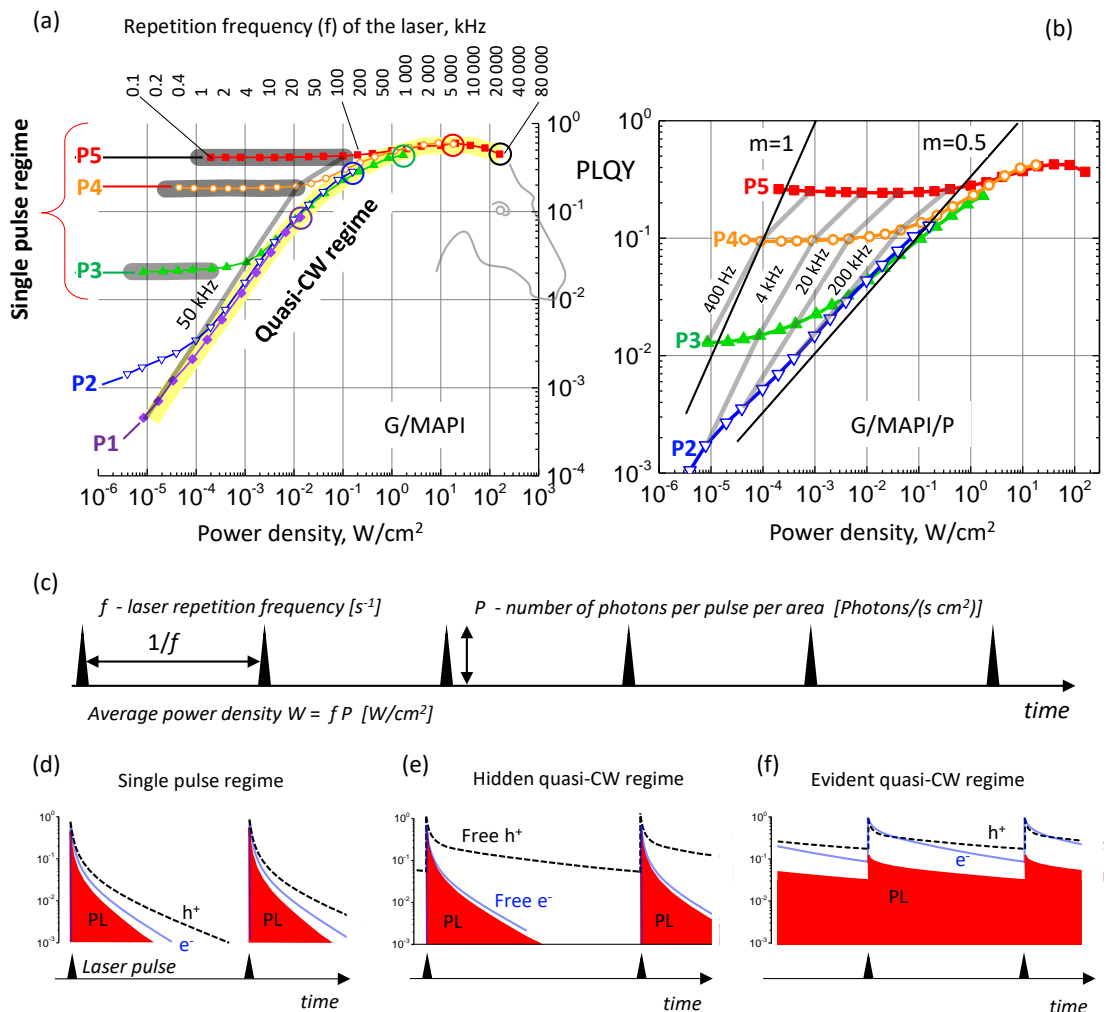

*Figure 1.* *General appearance of the PLQY(f,P) map and the details of the laser repetition rate (f) and pulse fluence (P) scanning. The pulse fluence was set one of the five values defined by the filters set placed in the laser beam:* ***P1**=4.1x10$^8$, **P2**=4.9x10$^9$, **P3**=5.1x10$^{10}$, **P4**=5.5x10$^{11}$ and **P5**=4.9x10$^{12}$ photons/cm$^2$*. In the single pulse regime these pulses create charge carrier densities of approximately 1.04x10$^{13}$, 1.24x10$^{14}$, 1.3x10$^{15}$, 1.37x10$^{16}$, 1.24x10$^{17}$ cm$^{-3}$ respectively. (a) PLQY(f,P) map for MAPbI$_3$ film grown on glass (G/MAPI) with the single pulse and quasi-CW excitation regimes indicated. Grey line shows PLQY(W) for f=50 kHz. (b) PLQY(f,P) map for MAPbI$_3$ film coated with PMMA (G/MAPI/P). Grey lines show the results of scanning of P at several fixed frequencies. The slope of these dependencies (m, PLQY ~ W$^m$ ) depends on the range of W and the value of f and can be anything from 1 to 0.5 (the corresponding dependencies W$^{0.5}$ and W$^1$ are shown by solid black lines). (c) – the excitation scheme. Illustrations of PL decays in the single pulse (d) and quazi-CW (e,f) excitation regimes. Here e$^-$ trapping is assumed leading to h$^+$ photodoping.*

The acquisition of a PLQY(f,P) map is fully automated (Supplementary Note 2) and includes precaution measures that minimize the exposure of the sample to light, while controlling for photo-brightening or darkening of the samples (Supplementary Note 4).[10,46,47] Such measures ensure that PLQY maps are fully reproducible when re-measured again on the same spot (see Supplementary Fig. S4.2). We note that the high degree of uniformity of our samples leads to very similar PLQY maps being measured on different areas of the sample (see Supplementary Notes 2 and 4).

Exemplary PLQY(f,P) maps for two different MAPbI$_3$ samples are presented in Fig. 1. The shape of the data point distribution resembles a "horse neck with a mane" and in the following, we will often employ this resemblance to refer to the different features of the PLQY(f,P) map. The data points for each value of *P* follow a characteristic line with a specific shape. When the frequency *f* exceeds a certain value, all data points start to follow a certain common dependency (the "horse neck"). The frequency at which this happens depends on P, such that, for example, the data obtained at pulse fluence P5 joins at *ca.* 200 kHz at the poll of the horse, while data collected at P1 joins at below 50 kHz, the withers of the horse (Fig. 1 a). The neck of the horse represents a regime in which PLQY depends solely on the averaged power density $W = f\, P\, h\nu$, where *hν* is the excitation photon energy.

For understanding the meaning of the PLQY(f,P) map we must distinguish between two principally different excitation regimes for a semiconductor:

1) *Single pulse regime:* in this regime the repetition rate of the laser is so low, that PLQY values and PL decays do not depend on the lag between consecutive laser pulses. In other words, the excited state population created by one pulse had enough time to decay to such a low level, that it does not influence the decay of the population generated by the next pulse (Fig. 1 d). In this case, PLQY does not depend on the lag between pulses (*i.e.* the pulse frequency). This regime is observed when PLQY follows the horizontal "hair strands" of the "horse mane" upon frequency scanning (highlighted in grey in Fig. 1 a).

2) *Quasi-continuous wave (quasi-CW) regime*: in this regime, the decay of the population generated by one pulse is dependent on the history of the excitations by previous pulses. This happens when some essential excited species did not decay completely during the lag time between the laser pulses (Fig. 1 e, f). In this regime, the data points follow the same trend and fall on the "horse neck", highlighted in yellow in Fig. 1 a.

The transition between the single pulse and quasi-CW regimes occurs when the individual "hair strands of the mane" (data points at fixed values of *P*) start to match with each other upon increasing *f* making the "neck of the horse". To conclude, PLQY(f,P) mapping allows for unambiguous and very easy discernment between the single pulse and quasi-CW excitation regimes.

However, the distinction between the single pulse and quasi-CW regimes is not at all obvious if the standard scanning over *P* is implemented at a fixed frequency *f* (see *e.g.* ref.[16,17]). To illustrate this, examples of the standard scans (grey lines) are shown in Fig. 1 b for several fixed values of *f*. We highlight this, since one may erroneously reason that by choosing a low enough frequency, it is possible to guarantee that the excitation is in the single pulse regime. However, as Fig. 1 b shows, the pulse frequency at which the quasi-CW regime changes to the single-pulse regime (the point when the "hair strand" splits off the "neck") depends on the pulse fluence *P*. The cause for this effect is the presence of a non-exponential decay of the excited state population as will be discussed in detail later. Thus, low pulse fluence may still result in a quasi-CW regime even for very low frequencies. In the example in Fig. 1 b for a pulse fluence P2, frequencies as low as 4 kHz still result in the quasi-CW regime, while, raising the pulse fluence by an order of magnitude (*i.e.* P3, P4, P5) brings the system to the single pulse regime at the same pulse repetition rate. We underscore that in order to identify the regime of excitation, one needs to scan the pulse frequency, rather than the pulse fluence.

## 3. PLQY(f,P) maps and PL decays(f,P) of polycrystalline MAPbI$_3$

Fig. 2 compares the PLQY(f,P) maps measured for MAPbI$_3$ films prepared with four different combinations of the interfaces (Fig. 2 f, Supplementary Note 5): MAPbI$_3$ deposited on glass (G/MAPI), MAPbI$_3$ deposited on PMMA coated glass (G/P/MAPI), MAPbI$_3$ deposited on glass and coated with PMMA (G/MAPI/P) and MAPbI$_3$ deposited on PMMA/glass and coated by PMMA from the top (G/P/MAPI/P). All samples exhibit the same PL and absorption spectra (Supplementary Note 6). Scanning electron microscopy (SEM) images (see Supplementary Note 6) show that all samples exhibit a very similar microstructure, which is not affected by the presence of PMMA layers. Despite all these similarities, the PLQY(f,P) maps are clearly different. To emphasise the differences, we added three horizontal lines that mark the PLQY at the single pulse regimes for the pulse fluences P3, P4 and P5 for the G/MAPI sample in Fig. 1 a. Additionally, black arrows were added to highlight the reduction in PLQY in certain regimes, which will be discussed in the following.

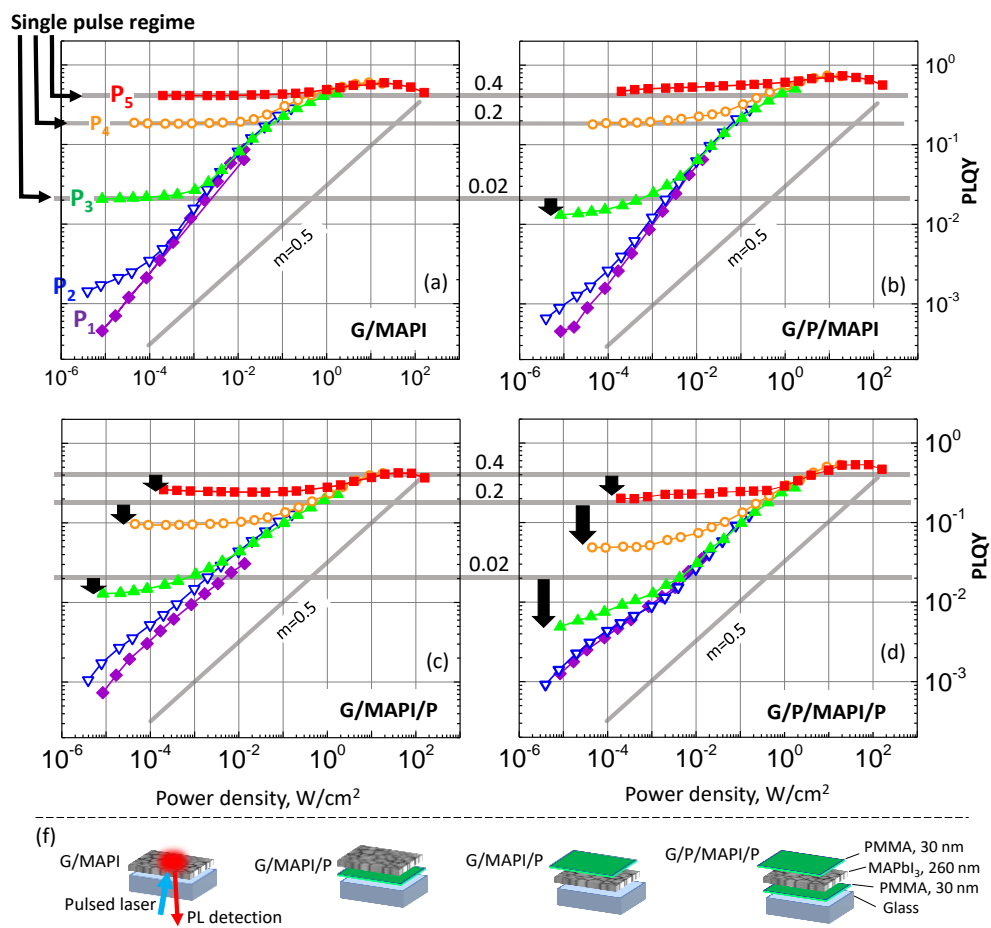

*Figure 2.* PLQY maps of the samples under study plotted in the same scale for comparison. (a) glass/MAPI, (b) glass/PMMA/MAPI, (c) glass/MAPI/PMMA, (d) glass/PMMA/MAPI/PMMA. The horizontal grey lines show the values of PLQY (0.4, 0.2 and 0.02) in the single pulse regime for the glass/MAPI sample (a) to set the benchmarks. Deviations from these values for other samples are shown by black arrows. The tilted grey line is the $W^{0.5}$ dependence as predicted by the SRH model. It is shown to see better the difference in the quasi-CW regime ("neck" of the horse) from sample to sample. The pulse fluence (P1 – P5) is indicted by the same colour code (shown in (a)) for all PLQY maps.

The decrease of PLQY upon the addition of PMMA differs for different values of *P*. Moreover, when comparing the slope m of the "horse neck" in (a) and (b) with that of (c) and (d), it is evident that it is strongly influenced by the exact sample stack. To visualise this difference, a line with the slope of m=0.5 (*i.e.* PLQY~$W^{0.5}$) is shown in each plot. The PLQY(f,P) map is most affected when MAPbI$_3$ film is coated by PMMA, while its presence at the interface with the glass substrate has only a minor effect.

Similar to the PLQY maps, PL decay kinetics also depends on the pulse fluence and excitation regime (single pulse *vs* quasi-CW). Such kinetics should be considered together with PLQY(f,P) map to complete the physical picture of charge recombination. Fig. 3 shows PL decays for the four types of samples as measured at f=100 kHz and pulse fluences P2 (low) and P5 (high). MAPbI$_3$ films deposited on glass (Glass/MAPI) exhibited the slowest of all PL decay kinetics both at a low and a high pulse fluences. The addition of PMMA to the sample stack accelerates the PL decay with the shortest decays observed for G/P/MAPI/P samples.

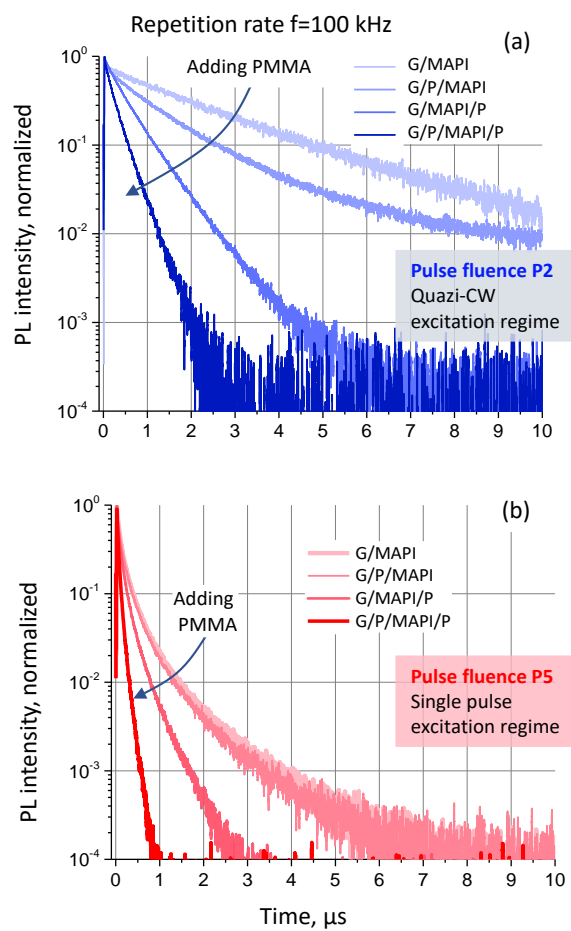

*Figure 3.* PL decays of all samples at 100 kHz repetition rate (10 µs distance between the laser pulses). (a) low pulse fluence (P2). (b) high pulse fluence (P5). Note that all decays in (a) are in the quasi-CW excitation regime, while all decays in (b) are in the single pulse excitation regime. Adding PMMA accelerates the PL decay.

The observation that modification of the sample interfaces by PMMA results in a faster PL decay not only for the low, but also for the high (P5) pulse fluence is particularly interesting. While the influence of surface modification on non-radiative recombination at low charge carrier concentrations is expected due to the changes in trapping, the same is not expected to occur at high pulse fluences. It is generally considered that at such fluences, the decays will be solely determined by non-linear processes such as Auger recombination and are thus not influenced by surface treatments. However, the change in decay dynamics in PMMA interfaced MAPbI$_3$ serves as the first indication that additional non-linear processes that involve trap states must be at play.

The second interesting observation is that according to the PLQY(f,P) map, the repetition frequency 100 kHz used for the PL decay measurements falls in the quasi-CW excitation regime for the low pulse fluence P2, but in the single pulse excitation regime for high pulse fluence P5. It is remarkable, however, that the PL intensity in the quasi-CW regime (Fig. 3 a) decays until the next laser pulse by almost two orders of magnitude for MAPbI$_3$ without PMMA and by four orders of magnitude for the sample coated with PMMA. This is an excellent example for the inability to correctly assign the excitation with P2 fluence to the quasi-CW excitation regime without the knowledge gained from the PLQY(f,P) maps, considering the population observable in the PL kinetics decays completely prior to the arrival of the next pulse. The cause for the quasi-CW regime in this case is the presence of a population of trapped carriers which lives much longer than 10 microseconds and that influences the dynamics via photodoping.[9,13,14] This example illustrates the 'hidden quasi-CW regime' shown schematically in Fig. 1e (see also Supplementary Note 7). These effects will be quantitatively explained by the theory detailed in the next section.

### 4. Theory and modelling

#### 4.1 Kinetic models: from ABC and SRH to SRH+

Fig. 4a schematically illustrates the key processes included in the ABC, SHR and extended SHR (SRH+) kinetic models. The SRH+ model contains terms for radiative (second order $k_r n p$) and non-radiative (all other terms) recombination of charge carriers. Non-radiative recombination occurs *via* a trap state or due to Auger recombination. The trapping process can be linear and quadratic (Auger trapping). Auger trapping refers to the process by which the trapping of a photoexcited electron provides an excess energy to an adjacent photoexcited hole.[2] The complete set of equations and additional description is provided in Supplementary Note 8. We note that in the SRH and SRH+ models, the complete set of equations for free and trapped charged carriers is solved, contrary to the studies where equations for only one of the charge carriers (*e.g.* electrons) are used (see ref. [43]). The latter simplification can work only if the concentration of holes is very large and constant (for example, in the case of chemical doping) which is not applicable for intrinsic MAPbI$_3$ and other perovskites. Due to the inclusion of Auger trapping in the SRH+ model, setting the parameter $k_n$ to infinity reduces it to the ABC model, where the coefficient *B* contains both radiative and non-radiative contributions. Finally, the SRH+ model reduces to the SRH model by ignoring all Auger processes.

Photon reabsorption and recycling are considered to be important processes that influencing the charge dynamics in MHPs.[11,44] In our experimental study we compare samples of very similar geometries and microstructure ensuring that the effects of photon reabsorption/recycling remain similar, such that they cannot serve as the reasons for the differences between PLQY(f,P) maps and PL decay kinetics amongst the different samples. As we discuss in detail in Supplementary Note 9, all effects on the charge dynamics related to the photon recycling in broad terms (both far field (photon reabsorption) and near field (energy transfer) effects), are included in our models via renormalized radiative rate and the Auger trapping coefficients, respectively. We also do not explicitly include charge diffusion in the model. The rational here is that charge carrier diffusion in MAPbI$_3$ occurs so fast that equilibrated homogeneous distribution of charge carriers over the thickness of the film can be assumed at a time scale of 10 ns and longer (Supplementary Note 8).

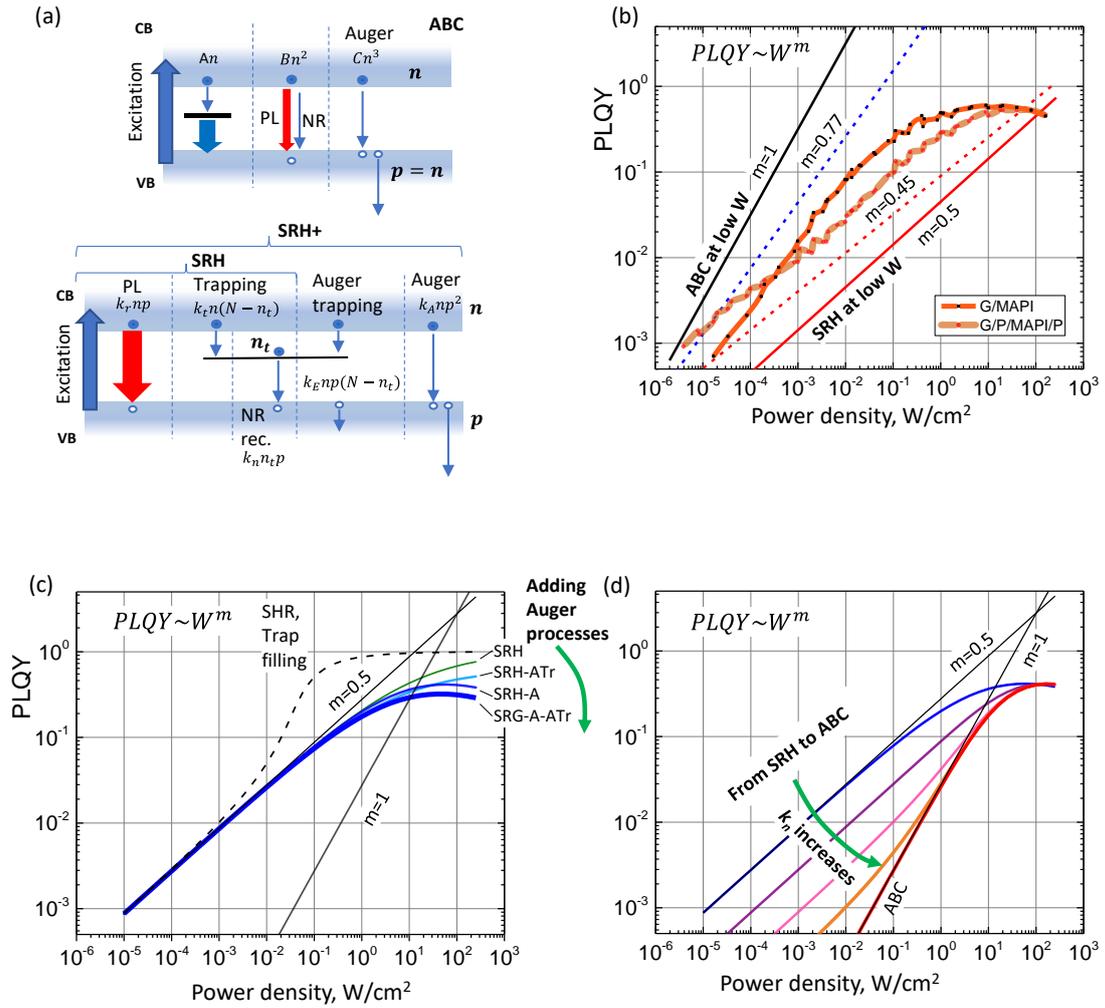

*Figure 4.* CW regimes of the ABC, SRH and SRH+ models and their comparison with the experiment. (a) The energy level scheme, the processes and parameters of all models (see the text and Supplementary Note 8 for details). (b) The experimental dependence (G/MAPI and G/P/MAPI/P samples) of PLQY on the excitation power density W in the quasi-CW excitation regime, m is the exponent in the dependence $W^m$. c) PLQY(W) in the CW regime for different models and trap feeling conditions. "-A" – adding Auger recombination, "-ATr" – adding Auger trapping (Supplementary Note 10). d) Evolution of the PLQY(W) upon transformation of the SRH model with Auger recombination to the ABC model with increasing of the parameter $k_n$. (see Supplementary Note 10 for the model parameters).

### 4.2. Applying the ABC, SRH and SRH+ models to the quasi-CW excitation regime

We first consider the CW excitation regime at low power densities. In this regime, the SHR and SRH+ models are identical since the contribution of Auger processes is largely negligible. Fig. 4 b shows the experimental dependencies of PLQY on the power density (*W*) for G/MAPI and G/P/MAPI/P samples and Fig. 4 c and d show the dependence calculated based on the three different models.

At low power densities PLQY(*W*) is a straight line in the double logarithmic scale (PLQY ~$W^m$) with the slope *m*=0.5 for the SRH and SRH+ models with no trap filling effect (see below) and *m*=1 for the ABC model.[1] Experimentally, we observe *m* ≈0.45 for those perovskite samples which are coated with PMMA (*e.g.* G/MAPI/P is shown in Fig. 4 b). This value is in a good agreement to the m=0.5 predicted by the SRH/SRH+ models in the case of the absence of trap filling. However, the other two samples, in which the $MAPbI_3$ surface is not coated with PMMA, exhibit *m*≈0.77 (*e.g.* G/MAPI sample is shown in Fig. 4b), which lies between the values of 1 and 0.5 predicted by the ABC and SRH/SRH+ models, respectively. These slopes are observed over at least four orders of magnitude in the excitation power density. Based on these results, we must conclude that $MAPbI_3$ samples with and without PMMA coating behave very differently in the quasi-CW regime.

In the framework of the SRH/SRH+ models, there are two possibilities that would lead to an increase in the coefficient *m*: (i) transformation toward the ABC model and (ii) trap filling effect in the SRH model. Fig. 4d shows the transformation of the SRH model, which includes Auger recombination to the ABC model by increasing the parameter $k_n$. At the condition $k_n \gg k_r, k_t$ there is a limited range of excitation power where one can obtain an intermediate slope *m* laying between 0.5 and 1 (Supplementary Note 8, Supplementary Note 10).

The second possibility is to allow for the trap filling effect to occur at the excitation power densities which are below the saturation of the PLQY due to the radiative and Auger recombination (the "horse head"). The effect of trap filling is caused when the number of available traps starts to decrease with increasing W. Consequently, the PLQY increases not only because the radiative process becomes faster (quadratic term), but also because the non-radiative recombination (trapping and further recombination) becomes smaller. As the result, PLQY grows faster than $W^{0.5}$ over a certain range of *W*. The effect is not trivial, because it is not the concentration of traps N as one would think, but rather the relation of $k_t$ to $k_r$ and $k_n$ (the necessary conditions is $k_t \gg k_r, k_n$), which determines if the trap-filling effect is observed in PLQY maps or not (Supplementary Note 8, Supplementary Note 10).

The trap filling effect is illustrated in Fig. 4 c, in which the parameter $k_t$ is increased whilst maintaining all other parameters fixed. Obviously, the resulting dependence is too strong and occurs over a too narrow range of excitation power densities (one order of magnitude) to fit the experimental data directly. Nevertheless, as will be shown below, such processes are present in $MAPbI_3$ samples which are not coated with PMMA, which exhibit a humped back of the "horse neck".

At high excitation densities, non-linear recombination processes begin to be particularly important. Since Auger processes are non-radiative, with further increase of W the PLQY cannot reach unity and instead decreases after reaching a certain maximum (the "horse head"). SRH cannot account for this effect considering it does not include any non-radiative non-linear terms and leads to PLQY=1 at high W. The ABC and SRH+ models can potentially describe this regime since they contain Auger recombination terms (Fig. 4 c and d).

### 4.3. Fitting of the PLQY(f,P) maps and PL decays kinetics by ABC, SRH and SRH+ models

To examine the validity of the three theories, we attempt to fit the experimental PLQY(f,P) plots and PL decays using all models and the results are shown in Fig. 5. Before we discuss the fitting results, it is important to stress that each simulated value of PLQY(f,P) at the PLQY maps and each PL decay curve shown in Fig. 5 are obtained from a periodic solution of the kinetic equations of the corresponding model under pulsed excitation with the required pulse fluence *P* and repetition frequency *f*. In practice it means that we excited the system again and again until the solution *PL(t)* stabilizes and begins to repeat itself after each pulse. Details of the simulations are provided in Supplementary Note 11.

When fitting experimental data, it is important to minimize the number of fitting parameters and maximize the number of parameters explicitly calculated from the experimental data. We exploit the experimental data to extract several parameters. First, considering that in all three models, the decay of PL at low pulse fluences is determined exclusively by linear trapping and is thus mono-exponential, we can extract the parameter $k_t N$ of the SRH and SRH+ models. Indeed, such behaviour is observed experimentally for the studied samples (see Fig. 3a) allowing us to use the decays at low pulse energies (P1-P3) to directly determine the trapping rates $k_t N$. We note, however, to obtain the best fit using the ABC model, the PL decays were not used to fix the parameter A. Secondly, in a single pulse excitation regime (*i.e.* the horizontal "mane of the horse"), the magnitudes of PLQY at pulse fluence P3 and P4 allow to determine the ratio $\frac{k_r}{k_t N}$ in SRH/SRH+ models and the ratio $\frac{k_r}{A}$ for the ABC model. Detailed block schemes of the fitting procedures are provided in Supplementary Note 11.

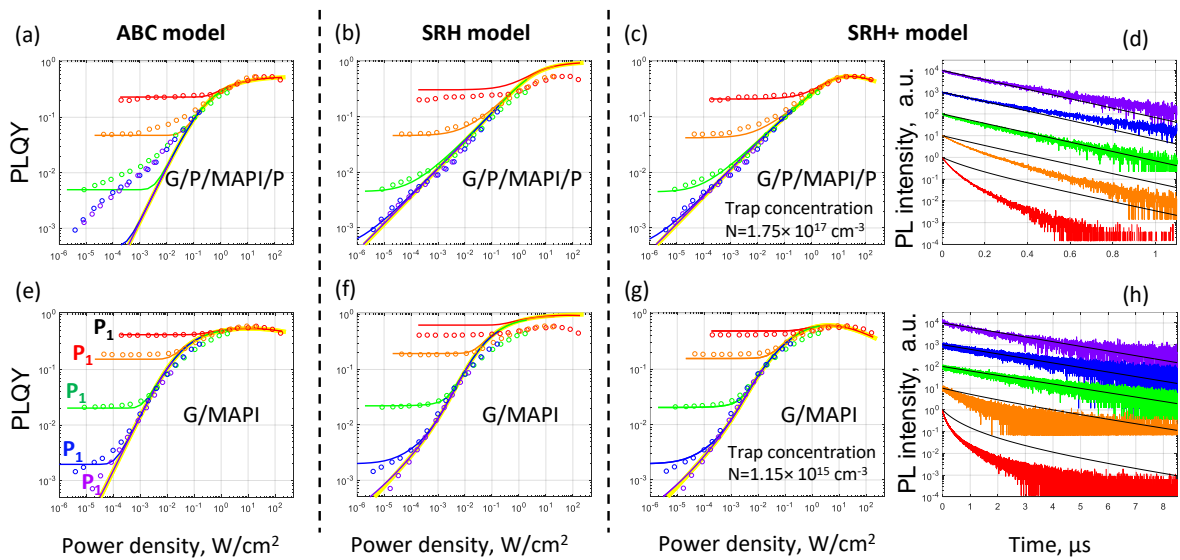

*Figure 5. Fitting of the PLQY(f,P) maps by all models. Bare MAPbI$_3$ film: (a) ABC, b) SRH, c) SRH+ and the MAPbI$_3$ film with PMMA interfaces:(e) ABC, (f) SRH, (g) SRH+. In PLQY maps the symbols are experimental points, the lines of the same colour are the theoretical curves. (d) and (h) show experimental and theoretical (black lines) PL decays according to the SRH+ model for both samples, laser repetition rate – 100 kHz. The pulse fluences are indicated according to the colour scheme shown in (e) in the whole figure. Theoretical CW regime is shown by the yellow lines in all PLQY maps. The model parameters can be found in Supplementary Note 12.*

As has been discussed above, MAPbI$_3$ samples coated with PMMA cannot be described using the ABC model due to mismatch of the slope within the quasi-CW regime (Fig. 5a), while both SRH and SRH+ models are well-suitable in this case (Fig. 5b, c). However, at a high excitation regime (*i.e.* the "head of the horse" in the quasi-CW and the single pulse regime at P5 pulse fluence) SRH+ works much better, highlighting the limitations of the SRH model on its own. Consequently, the entire PLQY(f,P) map of the PMMA coated films can be fitted using the SRH+ model with excellent agreement between the theoretical and experimental data (Fig. 5 c).

The behaviour of MAPbI$_3$ samples whose surface is left bare (exhibiting a "humped horse neck") can be approximated using the ABC model (Fig. 5e) and well-fitted by the SRH+ (Fig. 5 g) model. ABC indeed works quite well with, however, an obvious discrepancy in the tilt of the "neck". Very good fit can be obtained by the SRH/SRH+ models by adjusting of the $k_t$, $k_n$ and N to allow for the trap filling effect to occur in the medium excitation power range and, at the same time, making the dynamics closer to that in the ABC model by a relative increase of the recombination coefficient $k_n$ (see section 4.2 and Fig. 5g).

As was mentioned above, the PL decay rate at low power densities (P1-P3) was used to extract the product $k_t N$. This is the only occasion for which the PL decays are used in the fitting procedure of the SRH and SRH+ models. In the fitting procedure for the ABC model the PL decays are not used at all. Upon determining the fit parameters for each of the models, it is possible to calculate the PL decays at each condition and compare them with those decays measured experimentally. Importantly, PL decay rates calculated using the ABC model significantly underestimate the measured decay dynamics at all fluencies (Supplementary Note 12). On the other hand, as is shown in Fig. 5 d and h, the SRH+ model (as well as SRH, Supplementary Note 12) fit well the low fluence decay dynamics, but systematically underestimate the decay rate at high power fluences. It is noteworthy that the mismatch at the highest pulse fluence reaches a factor three for all samples, still significantly outperforming the fit using the ABC model. Insights regarding the applicability of the ABC, SRH and SRH+ models to the PLQY maps and PL decays are summarised in Table 1.

**Table 1**. Comparison of the ability of the three models to describe the PLQY and PL decays.

| Regime  Observables | Low and medium excitation pulse fluence power density (W< 0.1 Sun) | High excitation power density (1-300 Suns), high pulse fluence |
|---|---|---|
| **ABC model** | | |
| PLQY(f,P) | Quasi-CW regime - poor or very poor fit strongly depending on the sample  Good fit in the single pulse regime | Very good fit in all regimes |
| PL decays for given PLQY(f,P) | Cannot predict the PL decays | Cannot predict the PL decays |
| **SRH model** | | |
| PLQY(W,f) | Very good fit in all regimes | Discrepancy due to exclusion of high order processes |
| PL decays for given PLQY(f,P) | Very good match | Moderate underestimation of the initial decay dynamics |
| **SRH+ model** | | |
| PLQY(f,P) | Very good fit in both the quasi-CW and single pulse excitation regimes | Very good fit in both the quasi-CW and single pulse excitation regimes |
| PL decays for given PLQY(f,P) | Very good match | Moderate underestimation of the initial decay dynamics |

## 5. Discussion

### 5.1. The importance of pulse repetition frequency scanning in PLQY(f,P) mapping.

Scanning the excitation pulse repetition rate as proposed herein represents a novel experimental approach that transforms routine power dependent PLQY measurements to a universal methodology for elucidating charge carrier dynamics processes in semiconductors. In this PL-based method, we monitor not only the concentrations of free charge carriers, but also the concentration of trapped charges due to the total electro-neutrality of the system. Therefore, together with the time resolved PL decays, the PLQY map in the repetition frequency – pulse fluence parameter space comprise an experimental series which contains all the information concerning the charge dynamics in a given sample.

The most straight-forward application of the PLQY mapping is the unambiguous determination of the excitation regime of the experiment: single pulse *vs* quasi-CW. This is exemplified by the long-lived trapped charges ("dark" charges) in $MAPbI_3$ films that lead to a quasi-CW regime, which may seemingly contradict the decay of PL intensity by several orders of magnitude prior to the arrival of the next laser pulse (Fig. 3 a). Such trapped charges cause the so-called photodoping effect, which lingers until the millisecond timescale, and thus holds the "memory" of the previous laser pulse, leading to a stark influence on the PLQY map. While the importance of distinguishing between the single pulse and quasi-CW regimes has been noted in several publications before,[9,33] it has never been accomplished for MHPs experimentally. Indeed, in none of the published works presenting theoretical fits of experimental PLQY(W) dependencies this determination was possible simply because either only CW excitation[14,43] or pulse excitation with only one (625 kHz[17], 1 kHz[33]) or two (20 MHz and 250 kHz)[16] repetition rates of the laser pulses were employed.

To understand the excitation conditions regime is also critically important for interpretation of the classical experiments in which the PL intensity (or PLQY) is measured as a function of excitation power density (W) using a CW light source or a pulsed laser with a fixed repetition rate. Traditionally the intensity of PL is approximated using a $W^{m+1}$ dependence or in case PLQY is measured, with $W^m$ (because PLQY~PL/W), with both leading to a straight line in the double logarithmic scale.[1,13,33,35,45] According to the SRH and ABC models, approximations like this can be valid for a large range of W at low excitation power density only, when there is no trap filling effect, Auger processes can be neglected and PLQY is far from saturation. In all other cases, the dependence is not linear in the double logarithmic scale. As discussed above, SRH predicts *m*=0.5 in the CW excitation regime while ABC always predicts *m*=1. However, our experiments reveal that when the excitation is pulsed, one can obtain intermediate m values because upon increasing the power density, the experimental excitation regime is almost certainly switched from a quasi-CW to a single pulse (see grey lines in Fig. 1 a and b). Consequently, the extracted *m* not only doesn't fit to either model, but also cannot be reliably used for interpretation of the photophysics of the sample since any value of m can be obtained depending on the conditions of the pulsed excitation. Consequently, one needs to consider the excitation regime while attempting to link the slopes of PL/PLQY power dependencies to the photophysical processes taking place in the sample.

### 5.2. Applicability of the kinetic models

As we have shown above, neither the standard ABC nor the SRH model are capable of describing the complete PLQY maps and predicting PL decays of the investigated $MAPbI_3$ samples. On the other hand, the addition of Auger recombination and Auger trapping processes to the SRH model (SRH+ model) leads to an excellent fit of PLQY maps of all the studied samples. We emphasize that the (*f,P*) space used in this work is very large with $f$ varying from 100 Hz to 80 MHz (6 orders of magnitude) and pulse fluence $P$ changing over 4 orders of magnitude corresponding to charge carrier densities in the single pulse excitation regime from *ca.* $10^{13}$ to $10^{17}$ cm$^{-3}$. SRH+ model also agrees well with the PL decay kinetics for low and medium pulse energies (charge carrier concentrations from $10^{13}$ to $10^{15}$ cm$^{-3}$). However, for high pulse fluences ($10^{16}$ - $10^{17}$ cm$^{-3}$) the model underestimates the initial decay rate by up to a factor of three for the higher pulse energies, suggesting that SRH+ might also have certain limitations.

One possible explanation for the mismatch of decay rates at high excitation powers might be provided by considering experimental errors. It is well documented that the PL of perovskite samples is sensitive to both illumination and environmental conditions, which, may lead to both photodarkening or photobrightening of the sample.[9,46–48] To account for these effects, we paid a special attention to monitoring the evolution of the sample under light irradiation throughout the entire measurement sequence. As is shown in Supplementary Note 4, the maximum change in PL intensity during the entire measurement series is smaller than a factor of two. Taking this uncertainty together with other errors inherent to absolute PLQY and excitation power density measurements, missing the decay rates by a factor of three at the highest pulse fluence is not impossible. However, there is strong indication that the discrepancy reflects a problem of the model rather than in the experiment: the deviation between the theoretical and experimental PL decays is systematic. Experimental PL decay rates at high charge carrier concentrations are faster than predicted for all samples despite of the excellent matching of the PLQY(f,P) maps. A possible reason can be a presence of additional high-order recombination mechanisms calling for further development of the theory.

Despite of the moderate success at high charge concentration regime, the results of the SRH+ fitting still significantly outperform all previous attempts to explain charge carrier dynamics in MAPbI$_3$ samples and allow us to gain valuable insights concerning the photophysics of the samples investigated herein and the roles of traps within them. This is supported by the fact that the effect of charge trapping is the most crucial in the low and middle power ranges where the SRH+ model works very well for both the PLQY maps and PL decays.

The analysis of PLQY maps reveals that the concentration of dominant traps in high quality MAPbI$_3$ films (without PMMA coating) is approximately $1.2 \times 10^{15}$ cm$^{-3}$. This concentration is in excellent agreement with the range of values previously proposed by Stranks *et. al.*[14] where the trap concentration was estimated by assuming that PL decays become non-exponential exclusively due to the trap filling. We note, however, that trap filling is not a necessary condition to observe non-exponentiality in a PL decay. For that to occur, the non-linear recombination rate (radiative, Auger *etc.*) should just be faster than the trapping rate, which is determined not only by the trap concentration, but also by the capture coefficient. All these and related effects are considered when the data is modelled by the SRH+ model developed and employed here, thus allowing the extraction of the trap concentrations without any special assumptions.

Coating the top surface of MAPbI$_3$ with PMMA changes the picture drastically in terms of both the concentration and the nature of dominant traps. The concertation is increases by at least two orders of magnitude (N≥$1 \times 10^{17}$ cm$^{-3}$). More critically, the dominant traps in PMMA coated MAPbI$_3$ films exhibit a trapping efficiency and recombination rate per trap which are approximately one order of magnitude lower than the corresponding parameters for traps in uncoated MAPbI$_3$. It is these changes that lead to the absence of the visible trap filling effect in the PL dynamics. These results suggest that the addition of PMMA at the top surface leads to the creation of weak traps, which, however, due to their very large concentration override the effect of the stronger, yet less common, traps present in MAPbI$_3$ films that did not undergo a surface treatment. We note that PMMA coating is a common method employed in literature to protect MAPbI$_3$ samples from environmental effects when performing PL studies,[49,50] yet our results reveal that such a treatment fundamentally modifies the photophysics in the perovskite layer. More importantly, the supreme sensitivity of PLQY(f,P) mapping method to the influences of interfacial modifications illustrates its efficacy for studying charge carrier dynamics not only in films, but also in multilayers and complete photovoltaic devices.

**Conclusions**

To summarise, we examined the validity of the commonly employed ABC and SRH kinetic models in describing the charge dynamics of metal halide perovskite MAPbI$_3$ semiconductor. For this purpose, we developed a novel experimental methodology based on PL measurements (PLQY and time resolved decays) performed in the two-dimensional space of the excitation energy and the repetition frequency of the laser pulses. The measured PLQY maps allow for an unmistakable distinction between samples and more importantly, between the single-pulse and quasi-continuous excitation regimes.

We found that neither ABC nor SRH model can explain the complete PLQY maps for MAPbI$_3$ samples and predict the PL decays at the same time. Each model is valid only in a limited range of parameters, which may strongly vary between different samples. On the other hand, we show that the extension of the SRH model by the addition of Auger recombination and Auger trapping (SRH+ model) results in an excellent fit of the complete PLQY maps for all the studied samples. Nevertheless, even this extended model tends to systematically underestimate the PL decay rates at high pulse fluences pointing towards the existence of additional non-linear recombination processes in MAPbI$_3$.

Our study clearly shows that neither PL decay nor PLQY data alone are sufficient to elucidate the photophysical processes in perovskite semiconductors. Instead, a combined PLQY mapping and time-resolved PL decays should be used to elucidate the excitation dynamics and energy loss mechanisms in luminescent semiconductors.

**Acknowledgements.**

This work was supported by the Swedish Research Council (2016-04433) and Knut and Alice Wallenberg foundation (2016.0059). J.L. thanks China Scholarship Council (CSC No. 201608530162) for a PhD scholarship. Theoretical work was supported by the Russian Science foundation Project (20-12-00202). P.F. and S.S. thank the Wenner-Gren foundation for the visiting (GFOh2018-0020) and postdoctoral (UPD2019-0230) scholarships. This work was supported by the European Research Council (ERC) under the European Union's Horizon 2020 research and innovation program (ERC Grant Agreement No. 714067, ENERGYMAPS). Y.V. and Q.A. also thank the Deutsche Forschungsgemeinschaft (DFG) for funding the <PERFECT PVs> project (Grant No. 424216076) in the framework of SPP 2196. We thank Dr. Fabian Paulus for performing and analysing the XRD measurements and Prof. Jana Zaumseil for providing access to the XRD facilities.

**Methods**

**Thin Film Preparation.**

All samples were prepared from same perovskite precursor which was prepared with 1:3 molar ratio of lead acetate trihydrate and methylammonium iodide dissolving in dimethylformamide (Supplementary Note 5). For the samples with PMMA between the glass and perovskite layer, PMMA was spin-coated on the clean substrates with 3000 rpm for 30 s and annealed at 100 °C for 10 min. The perovskite precursor was spin-coated at 2000 rpm for 60 s on glass or glass/PMMA substrates, following by a 25 s dry air blowing, a 5 min room temperature drying and a 10 min 100°C annealing. For the samples with PMMA on top of the perovskite layer, no further annealing was applied after PMMA deposition.

**PL measurements.**

Photoluminescence microscopy measurements were carried out using a home-built wide-field microscope based on an inverted fluorescence microscope (Olympus IX-71) (Supplementary Note 1). We used 485 nm pulsed laser (ca. 150 ps pulse duration) driven by Sepria controller (PicoQuant) for excitation with repetition rate tuned from 100 Hz to 80 MHz. The laser irradiated the sample through an objective lens (Olympus 40X, NA = 0.6) with approximately 30 µm excitation spot size. The emission of the sample was collected by the same objective and detected by an EMCCD camera (Princeton Instruments, ProEM 512B). Two motorized neutral optical density (OD) filter wheels were used: one in the excitation beam path in order to vary the excitation fluence over 4 orders of magnitude and one in the emission path to avoid saturation of the EMCCD camera. The whole measurement of a PLQY(f,P) map was fully automatized and took approximately 3 hours (see Supplementary Note 2 for details). Automation was crucial for avoiding human errors in the measurements of so many data points (about 100 data points per "horse").

Time-resolved photoluminescence (TRPL) measurements were carried out using the same microscope, by adding a beam splitter in front of the EMCCD and redirecting a part of the emission light to a single photon counting detector (Picoquant PMA Hybrid-42) connected to a TCSPC module (Picoharp 300).

Absolute PLQY measurements were performed using a 150 mm Spectralon Integrating Sphere (Quanta-φ, Horiba) coupled through an optical fiber to a compact spectrometer (Thorlabs CCS200). Sample PL was excited by the same laser with 80MHz excitation repetition rate and 0.01 W/cm$^2$ excitation power density. This reference point was then used to calculate the absolute PLQY for all pulse fluences and frequency combinations (Supplementary Note 2, Supplementary Note 3).

It is important to stress that the whole acquisition of PLQY(f,P) was fully automatized and the sample was exposed to light only for the measurements. This led to a rather small total irradiation dose of about 200 J/cm$^2$ (equivalent to 2000 seconds of 1 Sun power) per one PLQY(f,P) map which accumulated over 85 acquisitions during about 4 hours for one PLQY map. Note, that 90% of this doze was accumulated with the maximum power P5 which gives 1600 Suns when the highest frequency 80 MHz is used. This allowed us to have minimal effects of light induced PL enhancement/bleaching on the measurements.